\documentclass[aps,prl,twocolumn,groupedaddress,nofootinbib]{revtex4}

\usepackage[utf8]{inputenc}
\usepackage{amsmath}
\usepackage{tikz}
\usepackage{extpfeil}
\usepackage{amsfonts}
\usepackage{amssymb} 
\usepackage{bm}
\usepackage{graphicx}
\usepackage{array}
\usepackage{lipsum}
\usepackage{float}
\usepackage{multirow}
\usepackage{hhline}
\usepackage{tabularx}
\usepackage{subfigure}
\usepackage{graphicx}
\usepackage{afterpage}
\usepackage{longtable}
\usepackage{epstopdf}

\newcommand{\cO}{\mathcal{O}}
\newcommand{\cM}{\mathcal{M}}
\newcommand{\mcs}{\cM_{cs}(X)}
\newcommand{\bP}{\mathbb{P}}
\newcommand{\bZ}{\mathbb{Z}}

\newcommand{\simpsec}[1]{\vspace{.5cm} \noindent \textbf{#1}}
\newcommand{\subs}[1]{\vspace{.5cm} \noindent \emph{#1}}

\begin{document}
\title{The Cost of Seven-brane Gauge Symmetry in a \\ Quadrillion F-theory Compactifications}
\author{James Halverson} 
\author{Jiahua Tian}
\affiliation{\vspace{.1cm}Department of Physics, Northeastern University, Boston, MA 02115-5000 USA} 


\begin{abstract}
  We study seven-branes in $O(10^{15})$ four-dimensional F-theory
  compactifications where seven-brane moduli must be tuned in order to
  achieve non-abelian gauge symmetry. The associated compact spaces
  $B$ are the set of all smooth weak Fano toric threefolds. By a
  study of fine star regular triangulations of three dimensional
  reflexive polytopes, the number of such spaces is estimated to be
  $5.8\times 10^{14}\lesssim N_\text{bases}\lesssim 1.8\times
  10^{17}$. Typically hundreds or thousands of moduli must be tuned to
  achieve symmetry for $h^{11}(B)<10$, but the average number drops
  sharply into the range $O(25)$-$O(200)$ as $h^{11}(B)$
  increases. For some low rank groups, such as $SU(2)$ and $SU(3)$,
  there exist examples where only a few moduli must be tuned in order
  to achieve seven-brane gauge symmetry.
\end{abstract}

\maketitle

\noindent \textbf{1. Introduction.}

Non-abelian gauge theories play a central role in particle physics, and a
celebrated question in string theory is whether such sectors should be expected. One might think the answer is
obvious, since they are implicit in certain ten-dimensional
superstring theories, but those symmetries can be broken by
compactification and restored only at special subloci in the
associated moduli space. Whether such enhanced symmetry loci exist,
the physics that stabilizes vacua on them, and the dynamics that might
drive the universe to those vacua are all relevant cosmological
questions in the string landscape; see
e.g. \cite{Kofman:2004yc,Grassi:2014zxa} and references therein.

F-theory \cite{Vafa:1996xn} is a generalization of the type IIb
superstring that allows the axiodilaton to vary in the extra spatial
dimensions. It provides a broad view of the landscape: in weakly
coupled type IIb limits \cite{Sen:1996vd,Sen:1997gv} it realizes the
best understood moduli stabilization scenarios
\cite{Kachru:2003aw,Balasubramanian:2005zx}, strong coupling effects
are computable by the power of holomorphy realized in complex
algebraic geometry \cite{Vafa:1996xn, MV}, and many landscape studies
have been performed in this context, see
e.g. \cite{Ashok:2003gk,Denef2004,Denef:2004cf,Denef:2006ad}.

Non-abelian gauge sectors may arise on seven-branes in F-theory, the
structure of which may be encoded in the geometry of a Calabi-Yau elliptic
fibration $X \xrightarrow{\pi} B$, where $B$ is the compact extra
dimensional space. In this paper we will focus on four-dimensional
compactifications, in which case $B$ is an algebraic threefold. The
seven-branes wrap a four-dimensional space $\Delta=0$ in $B$ that may have many
components, giving rise to many intersecting seven-branes. Each
seven-brane may carry an associated non-abelian gauge factor $G$ that
is determined in part by Kodaira's classification
\cite{MR0187255,MR0205280,MR0228019} of singular fibers, as well as
additional geometric data \cite{Bershadsky:1996nh} in $B$, T-branes
\cite{Cecotti:2010bp}, and G-flux \cite{Dasgupta:1999ss}. All are important, but the
geometrically determined data provides a foundation for the
seven-brane physics and is necessary for the existence of non-abelian
gauge symmetry, and therefore we focus on it here.  The associated
gauge group is more accurately called the geometric gauge group, but
for brevity we will drop such a distinction and refer to the gauge
group of a seven-brane. The gauge group of a seven-brane is determined by the structure of
$X$, but if the complex structure of $X$ is varied in the complex
structure moduli space $\mcs$, then the seven-branes may be deformed and
the gauge symmetry may be broken. The question of the existence of
loci with non-abelian gauge symmetry may then be studied in the
context of $\mcs$, which depends critically on the topology of $B$.

Recently there has been much work on so-called non-Higgsable clusters,
which are seven-brane gauge sectors that exist for generic points in
$\mcs$, and their properties are determined by $B$. These structures do not exist in eight dimensional
compactifications, but do in six
\cite{MV,Morrison:2012np,Morrison:2012js,Halverson:2016vwx,Morrison:2016lix}
and four
\cite{Anderson:2014gla,Grassi:2014zxa,Morrison:2014lca,Halverson:2015jua,Taylor:2015ppa}
dimensional compactifications.  Gauge factors that may appear on a
seven-brane in such a cluster include
\begin{equation}
G\in \{E_8, E_7, E_6, F_4, SO(8), SO(7), G_2, SU(3), SU(2)\}
\end{equation}
and the possible Lagrangian two-factor gauge sectors on adjacent
seven-branes in such a cluster are
\begin{align}
G_1\times G_2 \in \{SU(3)\times SU(3), G_2\times SU(2), \nonumber \\  SO(7)\times SU(2), SU(3)\times SU(2), \nonumber \\ SU(2)\times SU(2)\}.
\end{align}
Note that $SU(3)$ and $SU(2)$ are the allowed $SU(N)$ groups, whereas $SU(5)$ and $SO(10)$ never
occur for generic moduli \cite{Grassi:2014zxa}.
Non-Higgsable clusters in four-dimensional compactifications may have
interesting topologies \cite{Morrison:2014lca}, motivate
phenomenological models
\cite{Grassi:2014zxa,Halverson:2015jua,Taylor:2015ppa,Halverson:2016nfq},
and have implications for symmetry in the landscape
\cite{Grassi:2014zxa}.  The latter is strengthened by
analytical arguments and growing evidence
\cite{Morrison:2012js,Anderson:2014gla,Halverson:2015jua,Taylor:2015ppa} that
nearly all known extra dimensional spaces $B$ give rise to
non-Higgsable clusters.  

Conversely, some spaces $B$ do not exhibit seven-brane gauge symmetry
at generic points in $\mcs$. Seven-brane gauge symmetry often exists
on subloci in $\mcs$, though, in which case arriving in such a vacuum
requires that those vacua are stabilized and cosmologically
populated. While at this point it is difficult to address issues of
dynamics, recent estimates \cite{Braun:2014xka,Watari:2015ysa} show
that the number of flux vacua on subloci with symmetry is
exponentially suppressed relative to the number without symmetry (i.e. at generic points in $\mcs$). For example, there exist $B$ where
flux vacua with $SU(5)$ gauge group are suppressed
\cite{Braun:2014xka} by a factor of $e^{O(1000)}$ relative to those with no gauge symmetry. These suppression
factors get larger as the codimension in $\mcs$ necessary to obtain
gauge symmetry grows.

In this paper we study seven-brane gauge symmetry in a quadrillion, i.e. $O(10^{15})$,
four-dimensional F-theory compactifications that do not exhibit gauge symmetry at generic points
of $\mcs$. The results of \cite{Braun:2014xka} extrapolated to these compactifications
would imply that, for fixed $B$, the number of flux vacua with symmetry is exponentially suppressed
relative to the number of flux vacua without symmetry. Rather than computing numbers of flux vacua,
we will instead measure the ``cost'' of symmetry by computing the number of moduli that must be tuned to
engineer seven-brane gauge symmetry on any toric divisor in any smooth weak Fano toric threefold,
which are in one to one correspondence with fine star regular triangulations of the $4319$ reflexive
polytopes \cite{Kreuzer:1998vb}; a variety is weak Fano if $-K\cdot C > 0$ for all holomorphic curves $C$,
where $-K$ is an anticanonical divisor. We will do this for gauge groups in the set
\begin{align}
G\in \{&SU(2), SU(3),  SU(4), SU(5), \nonumber \\ &SO(7), Sp(1), Sp(2), SO(8), \nonumber \\ & SO(9), SO(10), G_2, F_4, E_6, E_7, E_8\},
\end{align}
some of which may arise in a number of ways.

If non-Higgsable clusters are a solution to a tuning (in moduli) problem in the
landscape, one goal of this paper is to diagnose the severity of the problem by studying models that do not exhibit non-Higgsable clusters. Compared to the results
of \cite{Braun:2014xka}, the larger set of spaces $B$ that we study suggests that the problem may be less 
severe. Specifically, spaces $B$ in this set with $h^{11}(B)<10$, which contain those of \cite{Braun:2014xka}, require tuning hundreds or
thousands or moduli in $\mcs$, but this number drops sharply into the range $O(20)$-$O(250)$ for $h^{11}(B)>20$.
It is reasonable to expect that the associated suppressions in ratios of flux vacua are much smaller than $e^{1000}$, though
likely still quite large. We leave vacuum statistics to future work.  We have also found examples where tuning a low rank group $G$ on a seven-brane on particular divisors
$D$ requires turning off only a few moduli; this is far from generic, but interesting nonetheless.

This paper is organized as follows. In Section $2$ we study fine-regular-star triangulations (FRST) of $3d$ reflexive
polytopes, which is important for statistical weighting of spaces $B$.  In Section $3$ we study the cost of tuning gauge symmetry on various types on
seven-branes wrapped on all toric divisors in all smooth weak Fano toric threefolds.
In Section $4$ we will conclude.

\newpage 
\simpsec{2. Landscape Estimates and Weights}

In this section we compute or estimate the number of fine-regular-star
triangulations (FRST) of each of the $4319$ $3d$ reflexive polytopes
\cite{Kreuzer:1998vb}. These are in one-to-one correspondence with
smooth weak Fano toric threefolds, which serve as the extra
dimensional spaces $B$ of the F-theory compactifications that we
study. 

Though the toric varieties associated to different FRST of the
same polytope have the same tuning costs, the number of FRST 
per polytope must be taken
into account in computing a weighted average of the cost of symmetry
across all $4319$ polytopes.
We will compute or estimate the number
of bases for each values of $h^{11}(B)$, which ranges from 1 to
35.\footnote{Note that there is no base space with
  $h^{11}(B)=33\ \text{or}\ 34$.} The details are described in the
following two subsections.

\subs{Approximate Number of FRST}

The number of FRST increases rapidly with $h^{11}(B)$, necessitating
different approximations for the number of FRST of a given polytope
as $h^{11}(B)$ increases. These methods will be called $A$, $B$, $C$,
$D$, and we will describe them in detail in this section.
Throughout, we the number of lattice points in $P$,
including the origin, as $n_P$. Hence, the corresponding toric variety has $h^{11}(B)=n_P-4$ and $n_P-1$ toric divisors. 

Method $A$ is to perform the exact calculation of the number of FRST.
Specifically, we compute the exact
numbers of FRST of the 1943 polytopes that have $n_P\leq14$, which
corresponds to $h^{11}(B)\leq10$. Computing all FRST for $h^{11}(B)=10$ takes multiple days of
computer time, motivating the use of approximation methods.

\begin{figure}[!htbp]
\centering
\includegraphics[width=0.5\textwidth]{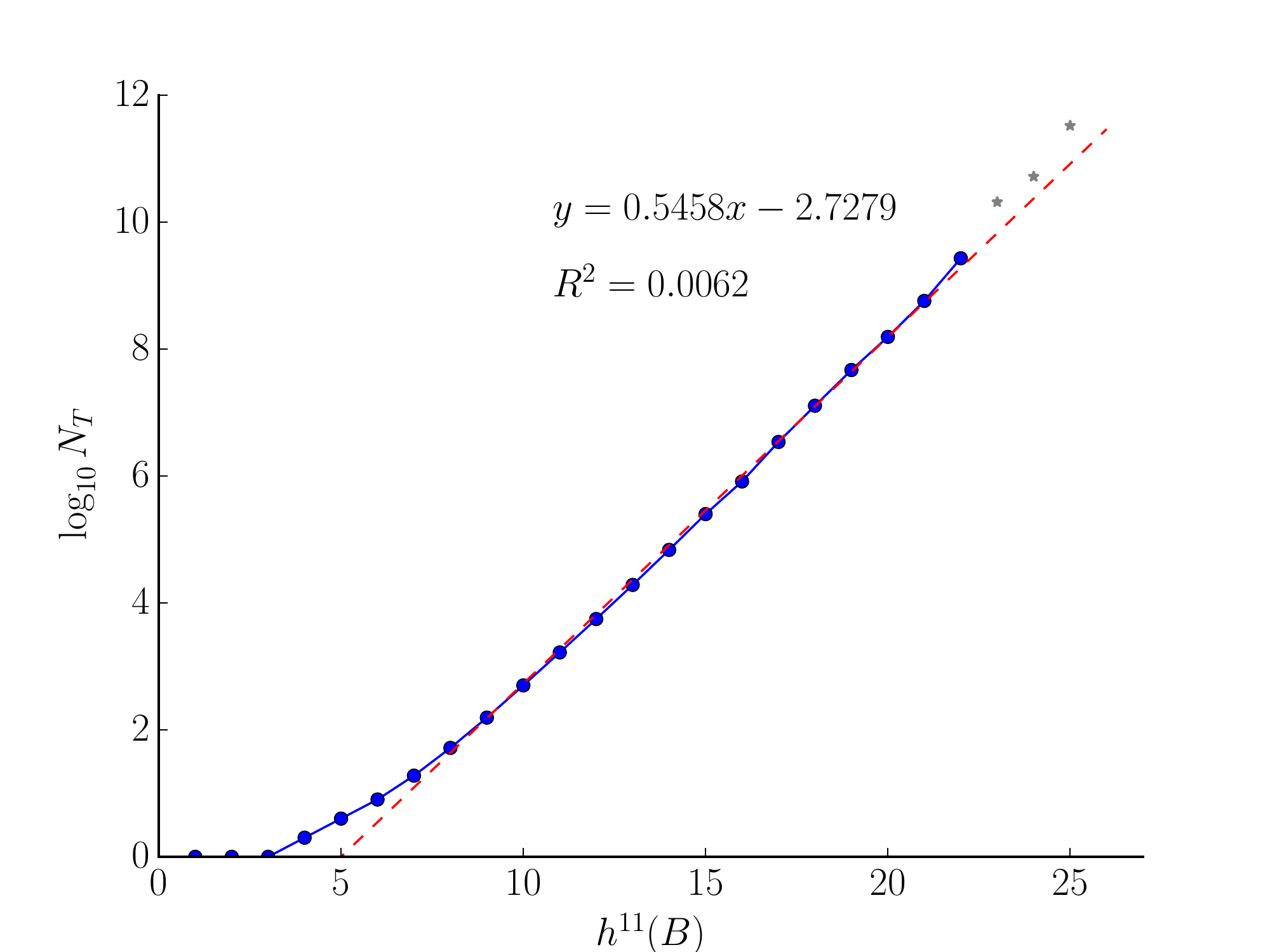}
\caption{The logarithm of the average numbers of triangulations $N_T$ per polytope when using different methods for $h^{11}_B\leq10$. The red curve denotes the exact result, i.e. method $A$, whereas the blue curve estimates
the number of FRST using method $B$.}\label{fig:Error}
\end{figure}

Method $B$ is our most accurate approximation method:
when $11\leq h^{11}(B)\leq22$ we
approximate the number of FRST of a polytope $P$ by the product of the
fine-and-regular triangulations (FRT) of each of its facets. This
approximation is justified by two facts: 1) This method
computes an estimated number of FRST within $10\%$ of the exact values for
$h^{11}(B)\leq10$, as shown in
Figure \ref{fig:Error}; 2) Only the order of
magnitude of number of FRST matters for our purposes, since this number can reach the order
$10^9-10^{17}$ when $h^{11}(B)\geq30$, and therefore a small error won't
qualitatively change the results.

\begin{figure}[!htbp]
\centering
\includegraphics[width=0.5\textwidth]{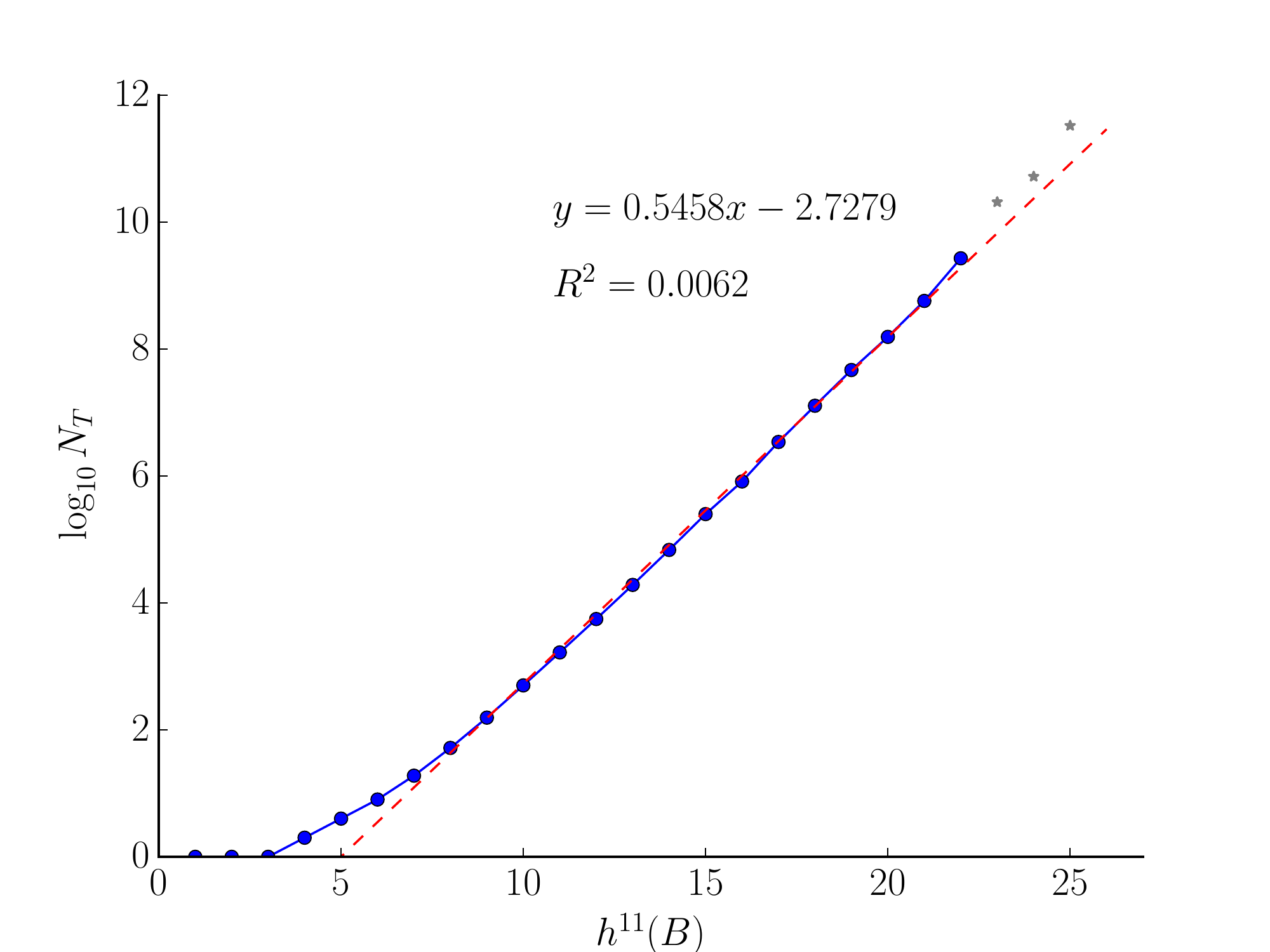}
\caption{The logarithm of the average number of triangulations $N_T$
  using method $B$. A line fits the data to high accuracy for $7\leq h^{11}(B)\leq22$. The grey stars shows on the plot are explained in the
 description of method $C$.}\label{fig:linefit}
\end{figure}

The results of method $B$ for $h^{11}(B)\leq22$ is shown in Figure \ref{fig:linefit}. Although we have computed the exact numbers of FRST for $h^{11}(B)\leq10$, the same estimate is also calculated for those cases so that we can see an approximately linear behavior of $\log_{10}N_T$ as a function of $h^{11}(B)$ when $7\leq h^{11}(B)\leq22$. 

We find that method $B$ is very time-consuming when applied to a polytope with $n_P\geq27$, which corresponds to $h^{11}(B)\geq23$. There are 91 polytopes with this property. Method $B$ breaks down for those 91 polytopes since triangulating the individual facets becomes too costly. That usually happens when the polytope contains a facet with $n_F\geq19$ where $n_F$ is the number of lattice points on the facet. 

To find a method to accurately estimate the number of FRT of such facets, we use the 
results of \cite{CLT} where the authors employed recursions that allows one to compute the number of triangulations for certain rectangular areas using dynamic programming. This allows us to obtain lower and upper bounds for the number of FRT of that facet, and then multiply with the exact numbers of FRT of the other facets to give an approximation to the number of FRST of the polytope. The calculation of these bounds on the number of FRT is highly dependent on the shape of the facet. 

We first focus on the 38 polytopes for which $n_P\geq30$ ($h^{11}(B)\geq26$). This relatively small set can be investigated case by case. The 8 polytopes shown in Table \ref{tab:NoEst} are those with $n_P\geq 30$ that have no facet\footnote{Actually there is a facet in polytope 296 with $n_F=19$, and this is the only case that the calculation can be done in a reasonable time. Since the others take too long to compute, we will just give the estimates for them.} with $n_F>19$.
Therefore the estimate can be done using
method $B$.
\begin{table}[!htbp]
\centering
\begin{tabular}{|c|c|c|c|c|c|}
\hline
$P$ & $h^{11}(B)$ & Number of FRST & $P$ & $h^{11}(B)$ & Number of FRST \\
\hline
2 & 31 & $3.034\times10^{15}$ & 4 & 26 & $1.275\times10^{12}$ \\
\hline
128 & 26 & $3.860\times10^{12}$ & 130 & 27 & $1.174\times10^{13}$ \\
\hline
134 & 26 & $1.809\times10^{12}$ & 136 & 26 & $2.773\times10^{12}$ \\
\hline
296 & 26 & $2.399\times10^{11}$ & 780 & 26 & $1.508\times10^{12}$ \\
\hline
\end{tabular}
\caption{These are the products of the numbers of FRST on each facet of the polytope. The polytope index is given by \textit{Sage 7.2} using \texttt{PALPreader}, indexing from one.}
\label{tab:NoEst}
\end{table}

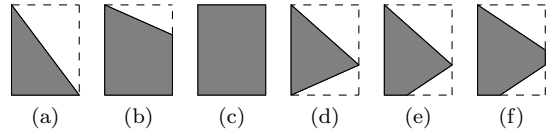
\begin{figure}[!htbp]
\centering
\subfigure[]{
\label{fig:Rcase}
\begin{tikzpicture}
\draw (0,0)--(0.9,0);
\draw (0,0)--(0,1.2);
\draw (0.9,0)--(0,1.2);
\draw [dashed] (0,1.2)--(0.9,1.2);
\draw [dashed] (0.9,1.2)--(0.9,0);
\draw[top color=black!50!white, bottom color=black!50!white] (0,0) -- (0.9,0) -- (0,1.2) -- cycle;
\end{tikzpicture}
}
\subfigure[]{
\label{fig:NRRcase}
\begin{tikzpicture}
\draw (0,0)--(0.9,0);
\draw (0,0)--(0,1.2);
\draw (0.9,0)--(0.9,0.8);
\draw (0,1.2)--(0.9,0.8);
\draw [dashed] (0,1.2)--(0.9,1.2);
\draw [dashed] (0.9,0.8)--(0.9,1.2);
\draw[top color=black!50!white, bottom color=black!50!white] (0,0) -- (0.9,0) -- (0.9,0.8) -- (0,1.2)-- cycle;
\end{tikzpicture}
}
\subfigure[]{
\begin{tikzpicture}
\label{fig:RECcase}
\draw (0,0)--(0.9,0);
\draw (0,1.2)--(0.9,1.2);
\draw (0.9,1.2)--(0.9,0);
\draw (0,0)--(0,1.2);
\draw[top color=black!50!white, bottom color=black!50!white] (0,0) -- (0.9,0) -- (0.9,1.2) -- (0,1.2) -- cycle;
\end{tikzpicture}
}
\subfigure[]{
\label{fig:NRcase}
\begin{tikzpicture}
\draw (0,0)--(0,1.2);
\draw (0,0)--(0.9,0.4);
\draw (0,1.2)--(0.9,0.4);
\draw [dashed] (0,0)--(0.9,0);
\draw [dashed] (0.9,0)--(0.9,0.4);
\draw [dashed] (0.9,0.4)--(0.9,1.2);
\draw [dashed] (0.9,1.2)--(0,1.2);
\draw[top color=black!50!white, bottom color=black!50!white] (0,0) -- (0,1.2) -- (0.9,0.4) -- cycle;
\end{tikzpicture}
}
\subfigure[]{
\label{fig:NRegcase}
\begin{tikzpicture}
\draw (0,0)--(0,1.2);
\draw (0.3,0)--(0.9,0.4);
\draw (0,0)--(0.3,0);
\draw (0,1.2)--(0.9,0.4);
\draw [dashed] (0.3,0)--(0.9,0);
\draw [dashed] (0.9,0)--(0.9,0.4);
\draw [dashed] (0.9,0.4)--(0.9,1.2);
\draw [dashed] (0.9,1.2)--(0,1.2);
\draw[top color=black!50!white, bottom color=black!50!white] (0,0) -- (0,1.2) -- (0.9,0.4) -- (0.3,0) -- cycle;
\end{tikzpicture}
}
\subfigure[]{
\label{fig:NNRegcase}
\begin{tikzpicture}
\draw (0,0)--(0,1.2);
\draw (0.3,0)--(0.9,0.4);
\draw (0,0)--(0.3,0);
\draw (0,1.2)--(0.9,0.6);
\draw (0.9,0.4)--(0.9,0.6);
\draw [dashed] (0.3,0)--(0.9,0);
\draw [dashed] (0.9,0)--(0.9,0.4);
\draw [dashed] (0.9,0.4)--(0.9,1.2);
\draw [dashed] (0.9,1.2)--(0,1.2);
\draw[top color=black!50!white, bottom color=black!50!white] (0,0) -- (0,1.2) -- (0.9,0.6) -- (0.9,0.4) -- (0.3,0) -- cycle;
\end{tikzpicture}
}
\caption{Shapes of facets}\label{fig:shapes}
\end{figure}

For $n_P\geq 30$ polytopes with a facet of $n_F\geq 19$ we compute
the bounds on the number of FRT triangulations. The computation depends on the
shape of such facets, which can be classified into 6 types as shown by the shaded areas in Figure \ref{fig:shapes}. The unshaded area is added to make the shape be rectangular so that the results in \cite{CLT} can be applied, as described previously.
The number of fine triangulations (FT) of these rectangular areas are exactly calculated in \cite{CLT} and at least $70\%$ of them are FRT;
This will suffice since again we are concerned with obtaining
the correct order of magnitude. The lower bound of the numbers of FT of the shaded area, denoted $N_{S}$, is determined by using the method of \cite{CLT} to calculate the numbers of FT of the strips of points
in this region, and then taking the product of them since gluing two fine-triangulated 2d point sets along a common edge gives a FT of their union. The upper bound is computed using the fact that $N_{S}\times N_{US}\leq N_{R}$, where $N_{US}$ is the number of FT of the unshaded area and $N_R$ is that of the total rectangular area.
Let $\kappa$ be the lower bound on $N_{US}$ computed by
the same method as for computing the lower bound of $N_S$. Then
$N_S\leq N_{R}/N_{US}\leq N_R/\kappa$ gives an upper bound on
$N_S$. 
We present the bounds on the number of FT of large facets of all polytopes with $n_P\geq 30$ in 
Table \ref{tab:AA}, and the associated bounds on the number of
FRST of those polytopes in Table \ref{tab:AAN}.
This is the approximation method we refer to as method $D$.
\begin{table}[!htbp]
\centering
\begin{tabular}{|c|c|c|c|}
\hline
$P$ & Case & Lower bound & Upper bound\\ 
\hline
8 & a & $6.112\times10^7$ & $4.471\times10^{10}$\\
\hline
10 & a & $9.459\times10^6$ & $2.415\times10^9$\\
\hline
12 & a & $6.236\times10^7$ & $3.919\times10^{10}$\\
\hline
18 & a & $4.204\times10^5$ & $6.434\times10^7$\\
\hline
74 & b & $2.037\times10^6$ & $1.550\times10^8$\\
\hline
80 & a & $1.323\times10^5$ & $1.614\times10^7$\\
\hline
88 & b & $8.732\times10^5$ & $2.080\times10^9$\\
\hline
90 & b & $6.306\times10^5$ & $4.099\times10^9$\\
\hline
100 & e & $6.418\times10^6$ & $7.026\times10^{11}$\\
\hline
223 & d & $3.209\times10^7$ & $2.616\times10^{11}$\\
\hline
236 & a & $1.323\times10^5$ & $1.614\times10^7$\\
\hline
256 & c & \multicolumn{2}{c|}{$7.370\times10^8$}\\
\hline
260 & b & $2.037\times10^7$ & $8.318\times10^9$\\
\hline
266 & b & $1.323\times10^5$ & $1.990\times10^8$\\
\hline
270 & c & \multicolumn{2}{c|}{$7.370\times10^8$}\\
\hline
272 & b & $9.459\times10^6$ & $7.307\times10^{9}$\\
\hline
293 & d & $4.204\times10^5$ & $3.666\times10^9$\\
\hline
348 & b & $9.460\times10^6$ & $7.307\times10^9$\\
\hline
352 & b & $6.306\times10^5$ & $4.657\times10^7$\\ 
\hline
387 & b & $1.715\times10^6$ & $1.842\times10^8$\\
\hline
521 & a & $4.204\times10^5$ & $6.434\times10^7$\\
\hline
527 & e & $6.306\times10^5$ & $6.200\times10^9$\\
\hline
553 & b & $6.305\times10^5$ & $2.990\times10^9$\\
\hline
762 & b & $1.774\times10^5$ & $4.657\times10^7$\\
\hline
798 & b & $1.323\times10^5$ & $1.990\times10^8$\\
\hline
810 & c & \multicolumn{2}{c|}{$7.370\times10^8$}\\
\hline
812 & b & $2.037\times10^6$ & $1.550\times10^8$\\
\hline
840 & f & $8.732\times10^5$ & $2.721\times10^8$\\
\hline
985 & d & $3.153\times10^6$ & $7.131\times10^9$\\
\hline
1000 & b & $1.774\times10^5$ & $2.666\times10^8$\\
\hline
\end{tabular}
\caption{Estimates of the number of FT for a facet}
\label{tab:AA}
\end{table}
\begin{table}[!htbp]
\centering
\begin{tabular}{|c|c|c|c|}
\hline
$P$ & $h_{11}$ & Lower bound & Upper bound\\ 
\hline
8 & 35 & $2.320\times10^{14}$ & $1.697\times10^{17}$\\
\hline
10 & 31 & $5.710\times10^{12}$ & $1.458\times10^{15}$\\
\hline
12 & 35 & $7.715\times10^{13}$ & $4.849\times10^{16}$\\
\hline
18 & 26 & $2.899\times10^9$ & $4.436\times10^{11}$\\
\hline
74 & 26 & $7.946\times10^{9}$ & $6.045\times10^{11}$\\
\hline
80 & 28 & $1.014\times10^{11}$ & $1.237\times10^{13}$\\
\hline
88 & 27 & $5.231\times10^{10}$ & $1.246\times10^{14}$\\
\hline
90 & 27 & $1.322\times10^{10}$ & $8.592\times10^{13}$\\
\hline
100 & 31 & $9.195\times10^{11}$ & $1.007\times10^{17}$\\
\hline
223 & 32 & $5.747\times10^{12}$ & $4.685\times10^{16}$\\
\hline
236 & 30 & $1.238\times10^{13}$ & $1.510\times10^{15}$\\
\hline
256 & 31 & \multicolumn{2}{c|}{$2.445\times10^{14}$}\\
\hline
260 & 31 & $7.323\times10^{12}$ & $2.990\times10^{15}$\\
\hline
266 & 27 & $2.076\times10^{11}$ & $3.122\times10^{14}$\\
\hline
270 & 27 & \multicolumn{2}{c|}{$1.192\times10^{15}$}\\
\hline
272 & 30 & $3.212\times10^{12}$ & $2.481\times10^{15}$\\
\hline
293 & 29 & $3.242\times10^{11}$ & $2.827\times10^{15}$\\
\hline
348 & 29 & $2.478\times10^{11}$ & $1.914\times10^{14}$\\
\hline
352 & 27 & $4.230\times10^{10}$ & $3.154\times10^{12}$\\
\hline
387 & 27 & $6.322\times10^{10}$ & $6.790\times10^{12}$\\
\hline
521 & 28 & $4.518\times10^{10}$ & $6.914\times10^{12}$\\
\hline
527 & 26 & $9.913\times10^{9}$ & $9.746\times10^{13}$\\
\hline
762 & 27 & $2.174\times10^{11}$ & $5.707\times10^{13}$\\
\hline
798 & 26 & $1.672\times10^{10}$ & $2.515\times10^{13}$\\
\hline
810 & 26 & \multicolumn{2}{c|}{$8.596\times10^{13}$}\\
\hline
553 & 27 & $1.057\times10^{10}$ & $5.014\times10^{13}$\\
\hline
812 & 28 & $4.119\times10^{11}$ & $3.134\times10^{13}$\\
\hline
840 & 26 & $2.942\times10^{10}$ & $9.169\times10^{12}$\\
\hline
985 & 28 & $6.609\times10^{10}$ & $1.495\times10^{14}$\\
\hline
1000 & 26 & $2.510\times10^{10}$ & $3.772\times10^{13}$\\
\hline
\end{tabular}
\caption{Approximation of the number of FRST for a given polytope by the product of the FRT on each facet. When there are more than 21 lattice points in a facet, we apply the estimate of the number of FRT for that facet that are shown in Table \ref{tab:AA}.}
\label{tab:AAN}
\end{table}

For $23\leq h^{11}(B)\leq25$ the situation is more difficult,
since application of method $B$ is generally time consuming and there are
too many polytopes to do a case by case analysis of bounds.
For some polytopes where there is no facet with $n_F\geq19$ 
it is possible to apply method $B$. Figure \ref{fig:linefit} depicts the average FRST estimates for these polytopes
obtained using method $B$, where the averages are
labelled by grey stars; note that they lie close to the linear fit line. For
these reasons we believe it is justified to take the average values of those cases as an approximation to the numbers of FRST for
polytopes with $23\leq h^{11}(B) \leq 25$; this is method $C$. 

\subs{Estimate of the number of base spaces}

In this subsection we estimate the number of base spaces that we consider,
which is equivalent to the number of FRST of $3d$ reflexive polytopes.
In order to get a more accurate estimate we take into account the effect of fan isomorphisms induced by lattice isomorphisms, which relate
identical toric bases $B$. Since there are no lattice isomorphisms between different polytopes, the associated fans cannot be isomorphic to each other,
and therefore only the fan automorphisms within a polytope need to be considered.

Fix a reflexive polytope. If a triangulation $\mathcal{T}_1$ can be brought to another triangulation $\mathcal{T}_2$ by a $GL(3,\mathbb{R})$ transformation of the lattice, $\mathcal{T}_1$ is equivalent to $\mathcal{T}_2$, and therefore the associated toric varieties are
identical, leading to an overcounting. This effect will be most
severe for those polytopes that give rise to the largest number
of bases, which occur for large $h^{11}(B)$. Since there are only 38 polytopes with $h^{11}(B)\geq26$, and these should account for the vast majority of bases,  we study potential overcounting in these examples on a case by case basis.
\begin{table}[!htbp]
\centering
\begin{tabular}{|c|c|c|c|}
\hline
$P$ & Cut-down by & $P$ & Cut-down by \\
\hline
2 & $4!$ & 4 & $2!\times2!$ \\
\hline
8 & $3!$ & 10 & $2!$ \\
\hline
12 & $2!$ & 74 & $2!$ \\
\hline
80 & $2!$ & 88 & $1!$ \\
\hline
128 & $3!\times2!$ & 130 & $2!$ \\
\hline
134 & $4!$ & 136 & $2!\times2!$ \\
\hline
223 & $2!$ & 236 & $3!$ \\
\hline
256 & $4!$ & 260 & $2!$ \\
\hline
266 & $2!$ & 270 & $2!\times2!$ \\
\hline
272 & $2!$ & 293 & $2!$ \\
\hline
296 & $2!$ & 348 & $2!$ \\
\hline
352 & $2!$ & 387 & $3!\times2!$ \\
\hline
521 & 1 & 553 & $2!$ \\
\hline
762 & $2!$ & 780 & $2!$ \\
\hline
810 & $2!$ & 812 & $2!$ \\
\hline
\end{tabular}
\caption{Cutting-down coefficients.}
\label{tab:CDE}
\end{table}
Note that if the facets have different numbers of lattice points then the FRT or FT of them can never be equivalent. For
these reason, the $8$ polytopes out of the $38$ that have
indices $18, 90, 100, 527, 798, 840, 985, 1000$ are free from
overcounting. 

If there are $K$ facets that can be brought to each other by a
$GL(3,\mathbb{R})$ transformation then the our estimate could be cut
down by at most a factor $K!$. This is because, for all the remaining
cases with a set $A_S$ of $K$ facets of the same number of lattice
points, these facets can be transformed into a configuration that they
are symmetric about a plane $H_p$ by an $SL(3,\mathbb{R})$ action which can be realized by applying several shear transformations successively. Therefore a triangulation of a facet
in $A_S$ can be taken to be equivalent to some triangulation of
another facet in $A_S$ by an $SL(3,\mathbb{R})$ transformation
followed by a reflection about $H_p$ together with a suitable
transformation of the facets not in $A_S$. This leads to a cut-down by
a factor $K!$ if such a transformation of the facets that are
not in $A_S$ always exists, and therefore $K!$ is the maximum
cut-down factor. The result is summarized in Table
\ref{tab:CDE}. From these results it can be seen that automorphisms
induce a reduction by at most a factor of $24$. After cutting down,
the estimate for lower and upper bounds on the number of FRST for
$h^{11}(B)\geq 26$ polytopes is given in Table \ref{tab:AANCut}.
\begin{table}[!htbp]
\begin{tabular}{|c|c|c|c|}
\hline
$h_{11}$ & $P$ & Lower bound & Upper bound\\ 
\hline
26 & 4 & \multicolumn{2}{c|}{$3.188\times10^{11}$}\\
\hline
26 & 18 & $2.899\times10^9$ & $4.436\times10^{11}$\\
\hline
26 & 74 & $3.973\times10^9$ & $2.023\times10^{11}$\\
\hline
26 & 128 &  \multicolumn{2}{c|}{$3.217\times10^{11}$}\\
\hline
26 & 134 & \multicolumn{2}{c|}{$7.538\times10^{10}$}\\
\hline
26 & 136 & \multicolumn{2}{c|}{$6.933\times10^{11}$}\\
\hline
26 & 296 & \multicolumn{2}{c|}{$1.120\times10^{11}$}\\
\hline
26 & 527 & $9.913\times10^{9}$ & $9.746\times10^{13}$\\
\hline
26 & 780 & \multicolumn{2}{c|}{$7.540\times10^{11}$}\\
\hline
26 & 798 & $1.672\times10^{10}$ & $2.515\times10^{13}$\\
\hline
26 & 810 & \multicolumn{2}{c|}{$4.298\times10^{13}$}\\
\hline
26 & 840 & $2.942\times10^{10}$ & $9.169\times10^{12}$\\
\hline
26 & 1000 & $2.510\times10^{10}$ & $3.772\times10^{13}$\\
\hline
27 & 88 & $5.231\times10^{10}$ & $1.246\times10^{14}$\\
\hline
27 & 90 & $1.322\times10^{10}$ & $8.592\times10^{13}$\\
\hline
27 & 130 & \multicolumn{2}{c|}{$5.87\times10^{12}$}\\
\hline
27 & 266 & $1.038\times10^{11}$ & $1.561\times10^{14}$\\
\hline
27 & 270 & \multicolumn{2}{c|}{$2.980\times10^{14}$}\\
\hline
27 & 352 & $2.115\times10^{10}$ & $1.577\times10^{12}$\\
\hline
27 & 387 & $5.268\times10^{9}$ & $5.658\times10^{11}$\\
\hline
27 & 553 & $5.285\times10^9$ & $2.507\times10^{13}$\\
\hline
27 & 762 & $1.087\times10^{11}$ & $2.854\times10^{13}$\\
\hline
28 & 80 & $5.070\times10^{10}$ & $6.185\times10^{12}$\\
\hline
28 & 521 & $2.259\times10^{10}$ & $3.457\times10^{12}$\\
\hline
28 & 812 & $2.060\times10^{11}$ & $1.567\times13^{13}$\\
\hline
28 & 985 & $6.609\times10^{10}$ & $1.495\times10^{14}$\\
\hline
29 & 293 & $1.621\times10^{11}$ &$1.414\times10^{15}$\\
\hline
29 & 348 & $1.239\times10^{11}$ & $9.570\times10^{13}$\\
\hline
30 & 236 & $2.063\times10^{12}$ & $2.517\times10^{14}$\\
\hline
30 & 272 & $1.606\times10^{12}$ & $1.241\times10^{15}$\\
\hline
31 & 2 & \multicolumn{2}{c|}{$1.264\times10^{14}$}\\
\hline
31 & 10 & $2.855\times10^{12}$ & $7.290\times10^{14}$\\
\hline
31 & 100 & $9.195\times10^{11}$ & $1.007\times10^{17}$\\
\hline
31 & 256 & \multicolumn{2}{c|}{$1.019\times10^{13}$}\\
\hline
31 & 260 & $3.662\times10^{12}$ & $1.495\times10^{15}$\\
\hline
32 & 223 & $2.874\times10^{12}$ & $2.343\times10^{16}$\\
\hline
35 & 8 & $3.867\times10^{13}$ & $2.828\times10^{16}$\\
\hline
35 & 12 & $3.858\times10^{13}$ & $2.425\times10^{16}$\\
\hline
\end{tabular}
\caption{Approximation of the number of FRST for the polytope by the product of the FRT on each facet after a proper cut-down.}
\label{tab:AANCut}
\end{table}

Taking these cut-downs into consideration and the fact that the numbers different bases for $h^{11}(B)\geq26$ are much larger than those for $h^{11}(B)\leq25$, their number of FRST provides a good
estimate of the total number of base spaces. 
Using those upper and lower bounds, we estimate that 
\begin{equation}5.780\times10^{14}\lesssim N_\text{bases} \lesssim 1.831\times10^{17}.\end{equation} Mathematically,
this is an estimate on the number of smooth weak Fano toric threefolds.

\simpsec{3. The Cost of Seven-brane Gauge Symmetry}

In this section we study the cost of tuning all gauge groups in Table
\ref{tab:Kodaira} on seven-branes wrapped on any toric divisor in
any smooth weak Fano toric threefold.

Recall from the introduction that geometric gauge symmetry on
seven-branes in F-theory can be encoded in the structure of a Calabi-Yau
elliptic fibration $X$.
Let $X\xrightarrow{\pi}B$ be this elliptic fibration with base $B$ given in
Weierstrass form 
\begin{equation}
y^2=x^3+fx+g,
\end{equation}
with associated discriminant locus
\begin{equation}
\Delta=4f^3+27g^2=0,
\end{equation}  
where $f\in \cO(-4K_B)$, $g\in\cO(-6K_B)$, and $K_B$ is the canonical
bundle of $B$. For generic $p\in B$, $\pi^{-1}(p)$ is a smooth
elliptic curve, and for a generic $p$ in $\Delta=0$, $\pi^{-1}(p)$ is
one of the singular fibers classified by Kodaira. The
set of Kodaira we consider is listed in Table \ref{tab:Kodaira}. The Kodaira fiber of
a particular component of the discriminant locus may be determined
from the order of vanishing of $f$, $g$, and $\Delta$ along that
component, and together with some additional data (see the appendix
for details) this determines the gauge symmetry of the
seven-brane on that component. 
\begin{table}[!htbp]
\begin{tabular}{|c|c|}
\hline
Fiber $F$ & Gauge group $G_F$ \\
\hline
$I_2$ & $SU(2)$ \\
\hline
$I_{3ns}$ & $Sp(1)$ \\
\hline
$I_{3s}$ & $SU(3)$ \\
\hline
$I_{4ns}$ & $Sp(2)$ \\
\hline
$I_{4s}$ & $SU(4)$ \\
\hline
$I_{5ns}$ & $Sp(2)$ \\
\hline
$I_{5s}$ & $SU(5)$ \\
\hline
$I_{1ns}^*$ & $SO(9)$ \\
\hline
$I_{1s}^*$ & $SO(10)$ \\
\hline
$II$ & -- \\
\hline
$III$ & $SU(2)$ \\
\hline
$IV_{ns}$ & $Sp(1)$ \\
\hline
$IV_{s}$ & $SU(3)$ \\
\hline
$I_{0ns}^*$ & $G_2$ \\
\hline
$I_{0s_1}^*$ & $SO(7)$ \\
\hline
$I_{0s_2}^*$ & $SO(8)$ \\
\hline
$IV^*_{ns}$ & $F_4$ \\
\hline
$IV^*_s$ & $E_6$ \\
\hline
$III^*$ & $E_7$ \\
\hline
$II^*$ & $E_8$ \\
\hline
\end{tabular}
\caption{The set of Kodaira's singular fibers that we study, together
with a label denoting whether or not they are split, and the associated
gauge group.}
\label{tab:Kodaira}
\end{table}

The bases $B$ that we study, as discussed in the last section, are
smooth weak Fano toric threefolds. The general seven-brane
configuration for such bases is one recombined seven-brane that does
not carry non-abelian gauge symmetry; mathematically, this means
that at a generic point in the complex structure moduli space of 
$X\to B$, for
any such $B$, the variety $\Delta=0$ is irreducible.

Obtaining non-abelian gauge symmetry, then, \emph{requires}
tuning in the complex structure of $X$ by tuning $f$ and $g$
such that $\Delta=0$ becomes reducible
\begin{equation}
\Delta = \prod_i \Delta_i,
\end{equation}
where each component $\Delta_i=0$ is a seven-brane that may carry
a different gauge symmetry according to the orders of vanishing
of 
$f$, $g$ and $\Delta$. A small deformation away from this sublocus
in $\mcs$ with symmetry gives a small Higgsing of the associated gauge theory,
the massive W-bosons of which are string junctions \cite{GaberdielZwiebach,DeWolfeZwiebach}
which can be systematically studied \cite{GrassiHalversonShaneson2013,Grassi:2014sda,Grassi:2014ffa} in the geometry.

Our goal is to measure the cost of this tuning by computing the
number of moduli that must be turned off in order to engineer Kodaira
fibers of certain types on certain divisors. Specifically, we perform
these computations for every Kodaira fiber in Table \ref{tab:Kodaira}, for every toric divisor in every smooth weak Fano toric
threefold. This computation is equivalent to computing the codimension in $\mcs$ on which a given Kodaira fiber type
exists along a given divisor.

These computations are carried out as follows. Fix a three-dimensional
reflexive polytope that contains integral points $v_i\in \bZ^3$, where $i$ is the index of the integral points. Let 
$P_n$ be the polytope whose points are in one-to-one correspondence
with global sections of $\cO(-nK_B)$; it is defined by
\begin{equation}
P_n = \{m\in \bZ^3\,|\,m\cdot v_i+n\geq0 \,\,\, \forall i\}.\label{eq:defpolar}
\end{equation}
To any $m\in P_n$ there is a monomial $\prod_i x_i^{m\cdot v_i + n}$
that has non-negative exponents by construction, and by this correspondence
we will henceforth
refer to such a monomial as a monomial in $P_n$. In the cases $n=4,6$
these monomials are the ones that may appear in $f,g$ respectively,
and if one constructs the most general $f$ and $g$ (that is, turns on
all monomials) then $\Delta=0$ is irreducible and there is no gauge
symmetry on seven-branes.  

Suppose we would like to compute how many monomials
must be turned off in order to engineer a Kodaira fiber $F$ and 
associated seven-brane gauge symmetry along along a given toric divisor 
$D_i:=\{x_i=0\}$. 
This can be done by sorting the monomials in $P_n$ 
according to the exponents (orders of
vanishing) of $x_i$. Then, using
the data computed in the appendix and this sorting, it is straightforward
to determine which monomials must
be turned off or back on in order to engineer a given Kodaira
fiber. Suppose $N_\text{off}$ monomials are turned off and $N_\text{on}$ are turned
on; it is always the case that $N_\text{off} > N_\text{on}$. Then, obtaining
a Kodaira fiber of type $F$ on $D_i$ requires tuning $N_\text{off}-N_\text{on}$
moduli, i.e. this gauge symmetry on a seven-brane
on $x_i=0$ occurs on a sublocus of codimension $N_\text{off}-N_\text{on}$ in $\mcs$.

For example, consider the case of a seven-brane on $D_i$ with
Kodaira fiber $I_2$, which corresponds to $SU(2)$ gauge symmetry
and is the F-theory lift of two coincident $D7$-branes. From the appendix, we see this occurs when
\begin{equation}
f_0=-3a_{20}^2, \,\,\, g_0=2a_{20}^2, \,\,\, g_1=a_{20}f_1,\label{eq:I2}
\end{equation}
where  $f_k\in\cO(-4K_{D_i}+(4-k)N_{D_i|B})$, $g_k\in\cO(-6K_{D_i}+(6-k)N_{D_i|B})$ and $a_{2k}\in\mathcal{O}(-2K_{D_i}+(2-k)N_{D_i|B})$. 
To engineer this form, the sections to be turned off are $f_0$, $g_0$ and $g_1$ while $a_{20}$ is to be turned on. To do so algorithmically, we construct $P_4$, $P_6$ and $P_2$ whose points are monomials in the most general $f$, $g$ and $a_2$. The subscript $k$ in $f_k$, $g_k$ and $a_{2k}$ indicates the exponents of $x_i$ in the monomials of the corresponding sections. In particular, in this case we need to find the points in $P_4$ that correspond to monomials that vanish to order 0, the points in $P_6$ that correspond to monomials vanish to order 0 and 1 and the points in $P_2$ that correspond to monomials vanish to order 0, all with respect to $x_i$. The sum of the numbers of such points in $P_4$ and $P_6$ is $N_\text{off}$ while the number of such points in $P_2$ is $N_\text{on}$ since in order to tune $I_2$ on $D_i$, all the $f_0$, $g_0$ and $g_1$ monomials have to be turned off except those of the form given by Equation \ref{eq:I2}. 

In this way, the cost of tuning a particular Kodaira fiber $F$ and associated gauge
symmetry $G_F$ on a toric divisor $D_i$ is $N_\text{off}-N_\text{on}$,
where $N_\text{off}$ and $N_\text{on}$ are computed in accordance with the sections that must be tuned
in order to achieve $F$ on $D_i$. We denote the cost of symmetry of tuning such $F$ on $D_i$ with associated
$3d$ reflexive polytope $P$ as $C(P,D_i,F)$. From this data we compute various statistics.

\begin{figure}[!htbp]
\centering
\includegraphics[width=0.55\textwidth]{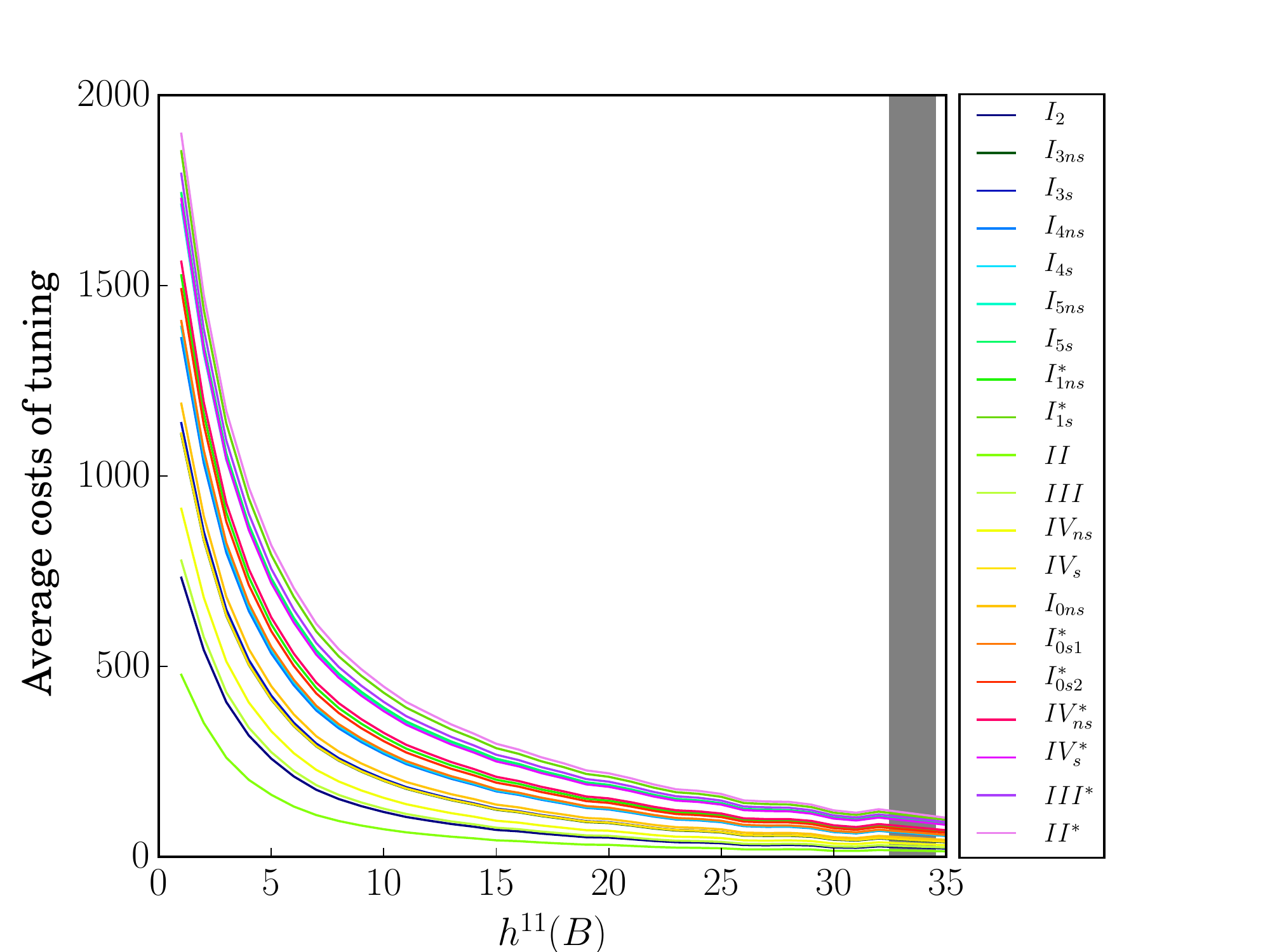}
\caption{Costs of unweighted tunings\textsuperscript{1}.}\label{fig:tuningall}
\end{figure}

We turn first to Figure \ref{fig:tuningall}, which is the plot of the unweighted average costs of tunings as a function of $h^{11}(B)$.
Let $S(h^{11}(B))$ be the set of $3d$ reflexive polytopes with $h^{11}(B)+4$ integral points, where the origin is included in the
set of integral points. 
Each line in the figure is a function of $h^{11}(B)$ and $F$, and the associated
unweighted average cost of symmetry for all toric divisors for all polytopes in $S(h^{11}(B))$ is
\begin{equation}
\frac{1}{|S(h^{11}(B))|}\sum_{P\in S(h^{11}(B))}  \sum_{D_i\in P} \frac{C(P,D_i,F)}{h^{11}(B)+3}.
\end{equation}
By unweighted, we mean that all polytopes are treated
equally, without taking into account the fact that they have different
numbers of FRST, and therefore different numbers of 
associated bases.  

There are two things that can be immediately read off from Figure
\ref{fig:tuningall}: 1) The unweighted average costs of tunings
decreases sharply as $h^{11}(B)$ increases, although there is small
fluctuation at the right end of the plot; 2) In general a gauge group
of a higher rank requires more tunings than those of a lower
rank. In particular, $II^*$ costs the most while $II$ costs
the least. The first point is due to the observation that a polytope with
higher $h^{11}(B)$ usually leads to polytopes  $P_n$ with less lattice
points, hence less monomials. The second point can be understood
by noting that tuning a gauge group of a higher rank usually
involves turning off more monomials in $P_4$ and $P_6$, which
dominates the overall costs since the monomials that are to be turned
on live in $P_2$, which has much less lattice points than $P_4$ and
$P_6$.  The reason we present the unweighted plot here is that it gives
a good approximation to the weighted plot, but can be understood by
simple arguments. Those arguments
can be made in the absence of the weighting process ``perturbing" the averaging, so that only the numbers of monomials
involved and the size of the polytopes $P_n$ matter.
\\
\footnotetext[1]{Shaded area indicates there are no varieties with $h^{11}(B)=32, 33$.}

\begin{figure}[!htbp]
\centering
\includegraphics[width=0.55\textwidth]{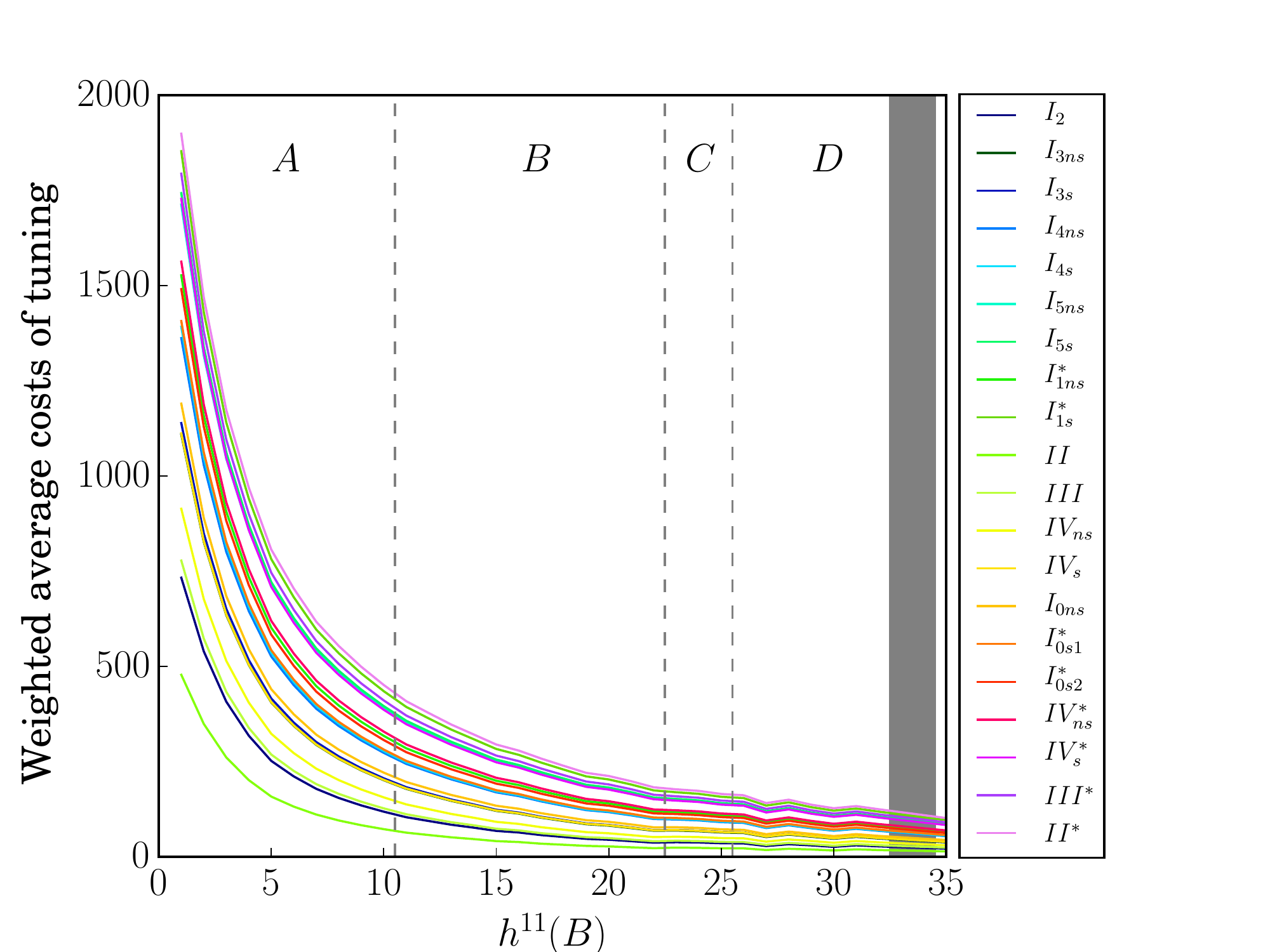}
\includegraphics[width=0.55\textwidth]{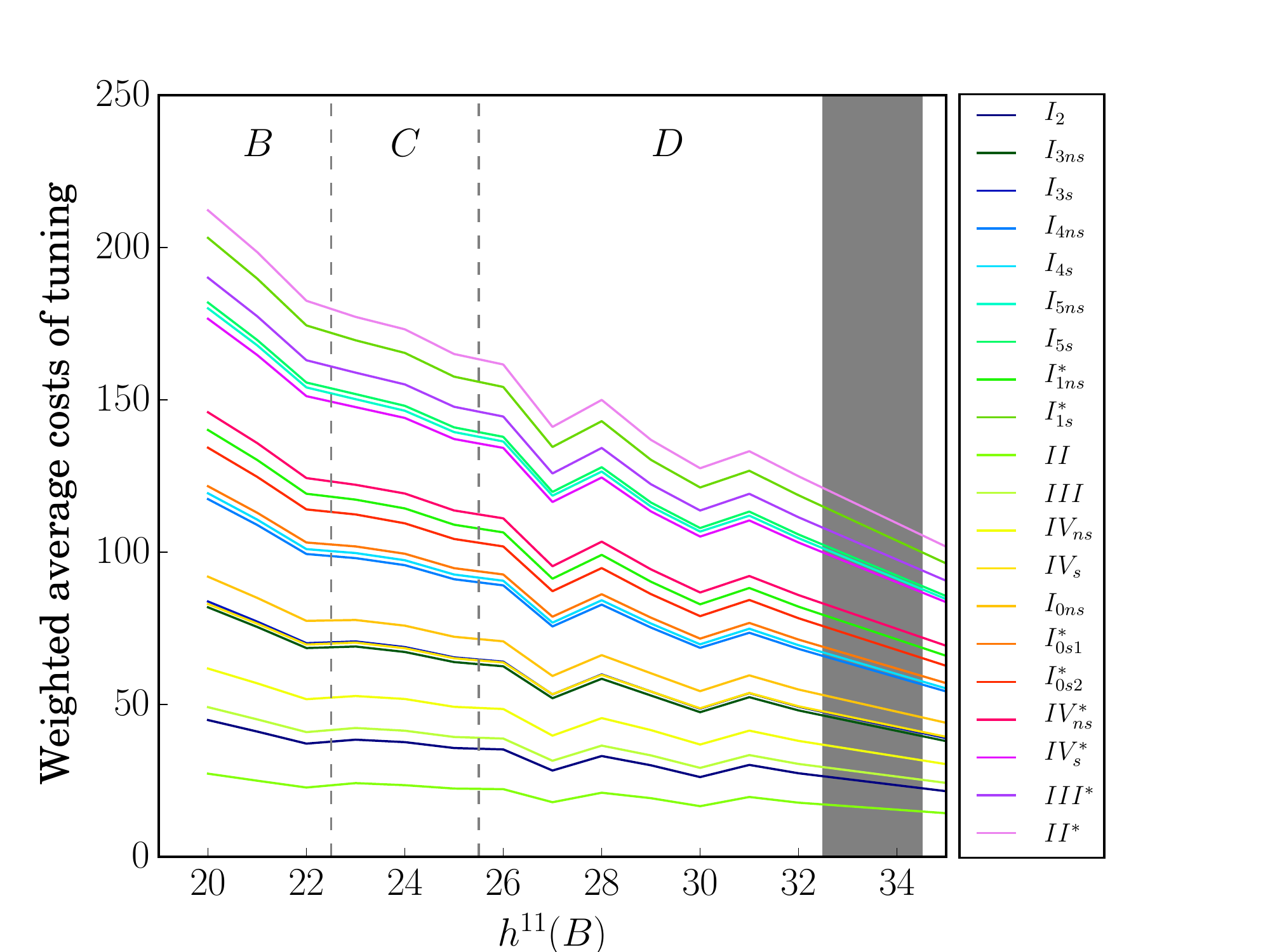}
\caption{Average costs of tunings after weighting by the numbers of FRST of the polytopes. The plot is divided into 4 regions. The labels $A$, $B$, $C$ and $D$ denotes the different methods we apply in the corresponding regions to get the numbers of FRST of the polytopes in the weighting process. \emph{Top:} Weighted average costs for all $h^{11}(B)$. \emph{Bottom:} Focusing on the region $h^{11}(B)\geq 20$.}\label{fig:weitot}
\end{figure}

For fixed $F$ and $G_F$, Table \ref{tab:maxandmin}
lists the maximal and minimal values of $C(P,D_i,F)$ for any toric divisors of a smooth weak Fano toric threefold.
Note that, contrary to common expectations, the maximal tunings do not occur for $B=\bP^3$ but instead for $B$ that are
associated to FRST of polytopes $7$ and $11$. Note also that though the average weighted costs of tuning in
Figure \ref{fig:weitot} is always at least $O(20)$, Table \ref{tab:maxandmin} shows that there do exist divisors where obtaining
non-abelian gauge symmetry only requires tuning a handful of moduli, for example a minimum of $2$ in the case of $SU(2)$.

\begin{table*}[!htbp]
\begin{tabular}{|c||c|c||c|c||c|}
\hline
Fiber & Maximum & Polytope & Minimum & Polytope & Gauge group\\
\hline
$I_2$ &1532 & 7, 11 & 2 & 8, 10, 16, 29, 35, 58, 88, 156, 178, 256, 258, 260, 387, 549, 670, 1462 & $SU(2)$ \\
\hline
$I_{3ns}$ &2226 & 7, 11 & 6 & 8, 10 & $Sp(1)$ \\
\hline
$I_{3s}$ &2289 & 7, 11 & 6 & 8, 10 & $SU(3)$ \\
\hline
$I_{4ns}$ &2632 & 7, 11 & 10 & 8 & $Sp(2)$ \\
\hline
$I_{4s}$ &2695 & 7, 11 & 10 & 8 & $SU(4)$ \\
\hline
$I_{5ns}$ &3147 & 7, 11 & 24 & 8 & $Sp(2)$ \\
\hline
$I_{5s}$ &3210 & 7, 11 & 24 & 8 & $SU(5)$ \\
\hline
$I_{1ns}^*$ &2913 & 7, 11 & 15 & 8 & $SO(9)$ \\
\hline
$I_{1s}^*$ &3364 & 7, 11 & 28 & 8 & $SO(10)$ \\
\hline
$III$ &1623 & 7, 11 & 3 & 8, 10, 16, 29, 35, 58, 88, 156, 178, 256, 258, 260, 387, 549, 670, 1462 & $SU(2)$ \\
\hline
$IV_{ns}$ & 1876 & 7, 11 & 4 & 8, 10, 16, 29, 35, 58, 88, 156, 178, 256, 258, 260, 387, 549, 670, 1462 & $Sp(1)$ \\
\hline
$IV_s$ & 2236 & 7, 11 & 7 & 8, 10 & $SU(3)$ \\
\hline
$I_{0ns}^*$ & 2372 & 7, 11 & 8 & 8 & $G_2$ \\
\hline
$I_{0s_1}^*$ & 2723 & 7, 11 & 11 & 8 & $SO(7)$ \\
\hline
$I_{0s_2}^*$ & 2858 & 7, 11 & 14 & 8 & $SO(8)$ \\
\hline
$IV_{ns}^*$ & 2968 & 7, 11 & 16 & 8 & $F_4$ \\
\hline
$IV_s^*$ & 3202 & 7, 11 & 22 & 8 & $E_6$ \\
\hline
$III^*$ & 3293 & 7, 11 & 26 & 8 & $E_7$ \\
\hline
$II^*$ & 3429 & 7, 11 & 30 & 8 & $E_8$ \\
\hline
\end{tabular}
\caption{Maximal and minimal costs of tuning for each fiber type and associated gauge group. Listed are the maximum and minimum values of $C(P,D_i,F)$ for given fibers and polytopes. Divisor index data is omitted in this plot.}
\label{tab:maxandmin}
\end{table*}

The plot of weighted average tuning costs is shown in Figure \ref{fig:weitot}, where the weighting takes into account
the estimated number of FRST per $3d$ reflexive polytope. For fixed $h^{11}(B)$ and $F$, this 
weighted average is defined by 
\begin{equation}
\sum_{P\in S(h^{11}(B))}\frac{N_P}{\sum_{P'\in S(h^{11}(B))} N_{P'}} \sum_{D_i\in P} \frac{C(P,D_i,F)}{h^{11}(B)+3}.
\end{equation}
Here the method of estimating the number of FRST $N_P$ depends on the value of $h^{11}(B)$. Recall from the
previous section that for $h^{11}(B)\leq10$, $N_P$ is the exact number of FRST of $P$. When $11\leq h^{11}(B)\leq22$ we apply method $B$ to estimate $N_P$, and when $23\leq h^{11}(B)\leq25$ we apply method $C$. Since applying method $C$ will assign the same $N_P$ to each $P$, in this region the weighted costs are numerically equivalent to the unweighted costs. When $h^{11}(B)\geq26$ we take $N_P$ to be the average of the lower bound and the upper bound given by method $D$. 

Quite surprisingly, comparing Figures \ref{fig:tuningall} and \ref{fig:weitot} we see the weighting does not significantly change the behavior of the costs of the tunings as a function of $h^{11}(B)$. The curves fluctuate a bit more in region $D$ but in regions $A$ and $B$, where the numbers of FRST of the polytopes are computed exactly or estimated more accurately, the percent difference between unweighted and weighted averages are no more than $16\%$. The maximal percent difference between the weighted and unweighted averages in any region is $21.5\%$. In regions $A$, $B$ and $C$ the cutting-down factor is not taken into consideration since a thorough investigation in these regions is very time consuming and
the large number of polytopes for lower $h^{11}(B)$ likely leads to an average cancellation effect. 
\\

\begin{figure}[!htbp]
\centering
\includegraphics[width=0.55\textwidth]{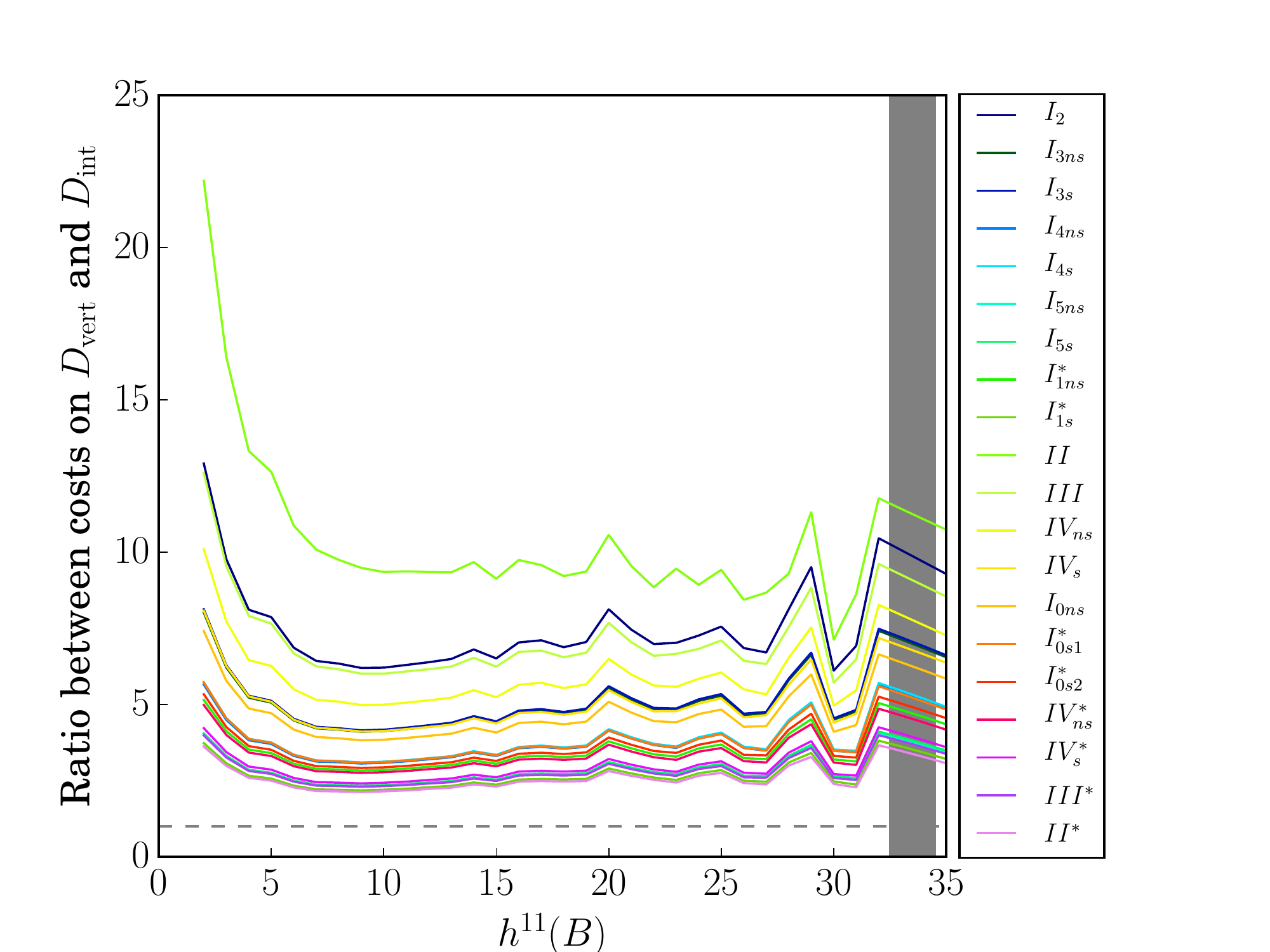}
\caption{Ratio between costs of tuning on $D_\text{vert}$ and $D_\text{int}$. The dashed line in the plot denotes $y=1$ which shows that the costs of tuning on $D_\text{vert}$ are always larger than those on $D_\text{int}$.}\label{fig:ratio}
\end{figure}

Our calculations show that tuning a fixed fiber type on a divisor $D_{\text{vert}}$ corresponding to a vertex of a $3d$ reflexive polytope $P$ 
usually costs more than on $D_{\text{int}}$ corresponding to a non-vertex lattice point of $P$. 
\begin{figure}[!htbp]
\centering
\includegraphics[width=0.55\textwidth]{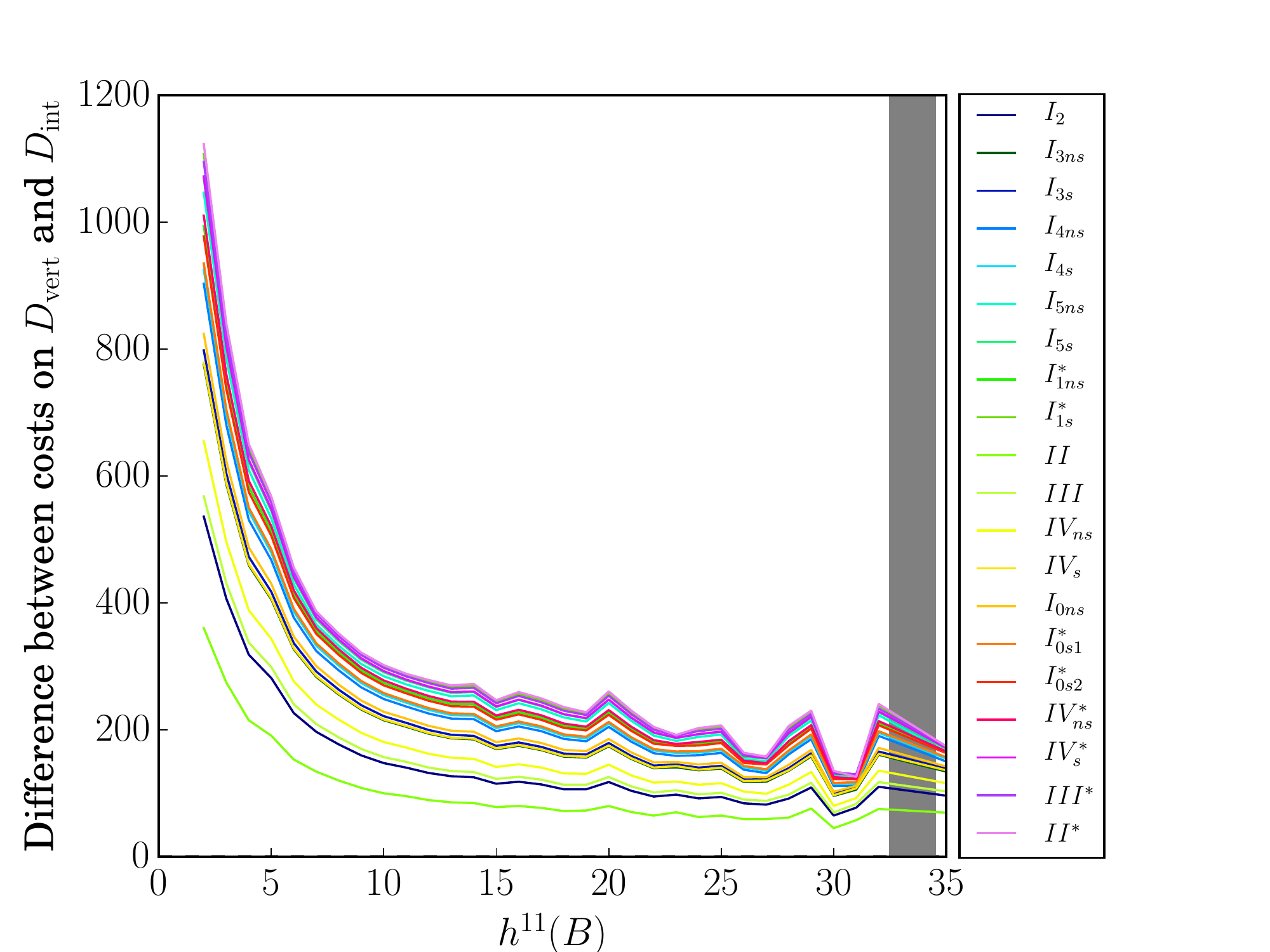}
\caption{Average difference between costs of tuning on $D_\text{vert}$ and $D_\text{int}$}\label{fig:difference}
\end{figure}
The average difference between the unweighted costs of tuning on $D_\text{vert}$ and $D_\text{int}$ is shown in Figure \ref{fig:difference}. Note that there are no non-vertex lattice points in $P$ when $h^{11}(B)=1$, hence there is no such difference. The ratio between average costs of tuning on $D_\text{vert}$ and $D_\text{int}$ is shown in Figure \ref{fig:ratio}.

A question naturally arises when we consider the Equation \ref{eq:defpolar}. Because of the linearity of this equation, it happens that tuning a gauge group on a divisor $D_1$ corresponding to $v_1$ may lead to tuning another gauge group on divisor $D_2$ corresponding to $v_2$. That happens when the monomials that are tuned to obtain a gauge group on $D_2$ is a subset of the monomials tuned to obtain a gauge group on $D_1$. A direct calculation shows that this never occurs when tuning fiber types $I_2, I_3, I_4$ and $I_5$. This can happen when tuning other fiber types,but any tuning on $D_1$ never forces the monomials $f$ and $g$ to vanish to order 4 and 6 on any other divisor respectively In one case that we have studied in depth, tuning $E_7$ on a divisor can lead to tuning other gauge groups on divisors that are adjacent
in some triangulations, but it never occurs that a higher rank gauge group arises in this process or it leads to $f$ and $g$ monomials that vanish to order 4 and 6 respectively on another divisor.

\simpsec{4. Discussion}

In this work we study the costs of tuning gauge groups on seven-branes in F-theory, where the
seven-brane is wrapped on any toric divisor in any smooth weak Fano toric threefold. These
may be constructed from FRST of 3d reflexive polytopes that are classified by Kreuzer and Skarke. A proper calculation of the weighted average costs requires an estimate of the numbers of FRST of the polytopes. The numbers of FRST of the polytopes with higher $h^{11}(B)$ dominate the number of base spaces so that an estimate of the numbers of FRST of those polytopes gives us an estimate of the total number of bases, $5.8\times 10^{14}\lesssim N_\text{bases}\lesssim 1.8\times10^{17}$. Thus, we study about
one quadrillion F-theory compactifications.

We investigate the polytopes $P_n$ in the dual lattice, whose points correspond to the sections of $\mathcal{O}(-nK_B)$ which can be turned off or on in order to tune a gauge group on a seven-brane on a divisor $D_i$. The sections involved are determined by the fiber type and can be sorted by their orders of vanishing along $D_i$. The exact numbers of the sections must be turned off and turned on to engineer
gauge symmetry can be calculated accordingly for each base space. This determines the codimension
of the sublocus in $\mcs$ on which seven-brane gauge symmetry exists on $D_i$; this is the
cost of symmetry. 

Both the unweighted and weighted average costs of symmetry are given in the paper. They are calculated according to Equation 10 and Equation 11 respectively, where our estimates of the number of FRST of the polytopes is applied in the weighting process. We find that both the unweighted costs and weighted costs drop steeply as $h^{11}(B)$ increases. When $h^{11}(B)\leq 5$ the average symmetry costs range from $O(250)$ to $O(2000)$; when $h^{11}(B)\geq10$ they drop down below $O(500)$; and when $h^{11}(B)>20$ they range from $O(25)$ to $O(200)$.

We find that the costs of tuning of the gauge groups depend significantly on the specific divisor on which symmetry is to be tuned. There exist divisors in certain polytopes on which only a few moduli need to be tuned to achieve symmetry, while there also exist divisors in certain polytopes where a few thousands moduli need to be tuned. The average costs as a function of $h^{11}(B)$ are plotted in Figures \ref{fig:tuningall} and \ref{fig:weitot}. We also show that cost of symmetry depends on the type of divisor. Specifically, tuning symmetry on divisors that correspond to a vertex of the 3d reflexive polytope costs more than tuning on the divisors that correspond to the interior points. 

In summary, our study of F-theory compactifications without non-Higgsable clusters shows that when non-abelian gauge symmetry must be tuned, it typically occurs on subloci in $\mcs$ of codimension $O(25)$-$O(250)$. While this cost of symmetry is less severe than in the work of \cite{Braun:2014xka}, it still
represents a significant cosmological challenge. We believe this result
further motivates the study of moduli stabilization on special subloci in moduli space, and also the study of non-Higgsable clusters.

\vspace{.5cm}
\noindent \textbf{Acknowledgements.}  We would like to thank Ross
Altman, Andreas Braun, Jonathan Carifio, Cody Long, Brent Nelson, Ben
Sung, and Washington Taylor for discussions about this work.  JH is
supported by NSF Grant PHY-1620526 and startup funding from
Northeastern University. He thanks the Center for Theoretical Physics
at MIT for hospitality at various stages of this work. This work was
performed in part at the Aspen Center for Physics, which is supported
by National Science Foundation grant PHY-1066293.

\section{Appendix}

Let $X\xrightarrow{\pi} B$ be an elliptic fibration in Weierstrass
form, where $y^2=x^3+fx+g$ with $f\in \cO(-4K_B)$ and $g\in
\cO(-6K_B)$, which has a discriminant locus $\Delta = 4f^3 + 27g^2$. We will consider the tuning of gauge symmetry on a divisor $Z$
defined locally by $z=0$; in toric cases, $z$ will be a homogeneous coordinate and therefore
$Z$ a toric divisor. This allows us to expand $f$ and $g$ as
$f=f_0+f_1z+f_2z^2+f_3z^3+\dots$ and $g=g_0+g_1z+g_2z^2+g_3z^3+\dots$,
where $f_i\in\cO(-4K_Z+(4-i)N_{Z|B})$ and
$g_i\in\cO(-6K_Z+(6-i)N_{Z|B}).$ We will also utilize
$a_{ni}\in\cO(-nK_Z+(n-i)N_{Z|B})$, which will be relevant in studying tuning.  Since we are only interested in
the structure of gauge groups, we will only consider codimension one singularities.
Throughout we use ``node'' to mean the node of a Dynkin diagram, represented geometrically by a $\bP^1$ fiber. Relevant arguments of the conditions under which the intersection diagrams of the nodes are the duals of the Dynkin diagrams of the gauge groups can be found in \cite{Katz:2011qp,Taylor:2015ppa}.

We are now ready to investigate the singularities of the elliptic fibration. We will study the relationship between
sections the ensures the existence of each fiber type, and in some cases we will resolve the codimension one singularities
to study their structure. We denote resolutions by notation of the form $X\xleftarrow{(x_1,x_2,\dots|e)}X'$ where the resolution
is performed along $\cap_i \{x_i=0\}$ and $e=0$ is the exceptional divisor; this gives replacements $x_i\mapsto ex_i$. 
The required tunings are summarized in Table \ref{appendixtable1}. 
\vspace{.5cm}

$\mathbf{I_1}$. If $X$ is smooth, then $\Delta$ has generic Kodaira fiber $I_1$. An $I_1$ fiber along $Z$ can be obtained
by $f_0=-3a_{20}^2$ and $g=2a_{20}^3$. $X$ is smooth.

$\mathbf{I_2.}$ Let $f_0=-3a_{20}^2, g=2a_{20}^3,
g_1=-a_{20}f_1$. There is one singularity at $x=a_{20}, y=0,
z=0$. Shift the singularity to $(0,0,0)$ by redefining $x$ and do the
blow up $X\xleftarrow{(x,y,z|e)}X'$. There are no more codimension 1
singularities after the blow-up. The exceptional divisor $e_0$ is
fibered by genus $0$ curves in the new $\mathbb{P}^2$ associated with
the blow-up. This gives us $G=SU(2)$.

$\mathbf{I_3.}$ Let $f_0=-3a_{20}^2, g_0=2a_{20}^3, g_1=-a_{20}f_1,
f_1=a_{20}a_{21}, g_2=\frac{a_{20}}{12}(a_{21}^2-12f_2)$. There is one
singularity at $x=a_{20}, y=0, z=0$. Shift it to $(0,0,0)$ and do the
blow up $X\xleftarrow{(x,y,z|e)}X'$. There are no more codimension 1
singularities after the blow-up. An analysis of $X'$ shows that $e=0$
is fibered by two lines that swap under encircling $a_{20}=0$ if and
only if $a_{20}$ is not a perfect square.  Thus, if $a_{20}$ is a perfect
square, $G=SU(3)$; if it is not, $G=Sp(1)=SU(2)$. 

$\mathbf{I_4.}$ Let $f_0=-3a_{20}^2, g=2a_{20}^3, g_1=-a_{20}f_1,
f_1=a_{20}a_{21}, g_2=\frac{a_{20}}{12}(a_{21}^2-12f_2),
g_3=\frac{1}{216}(a_{21}^3+36a_{21}f_2-216f_3)$.  Again we shift the
singularity to $(0,0,0)$ and do the blow-up
$X\xleftarrow{(x,y,z|e_1)}X'$. The resulting variety is still
singular. The singular locus is along $e_1=0, x=-\frac{a_{21}z}{6},
y=0$ so we shift it to $(0,0,0)$ and do a second blow up
$X'\xleftarrow{(x,y,e_1|e_2)}X''$. There are no more codimension 1
singularities after the blow-up. The fiber of $e_1=0$ splits into two
lines that swap upon encircling $a_{20}=0$ if and only if $a_{20}$ is
not a perfect square. Henceforth by ``split'' we will mean that two
nodes do not swap under encircling some locus, here $a_{20}=0$.  A
further analysis of $X''$ fibers shows that $G=SU(4)$ if $a_{20}$ is a
perfect square then $G=SU(4)$; if not, then $G=Sp(2)$.

$\mathbf{I_5.}$ Let $f_0=-3a_{20}^2, g=2a_{20}^3, g_1=-a_{20}f_1,
f_1=a_{20}a_{21}, g_2=\frac{a_{20}}{12}(a_{21}^2-12f_2),
g_3=\frac{1}{216}(a_{21}^3+36a_{21}f_2-216f_3),
g4=-a_{20}f_4+\frac{a_{21}f_3}{6}, f_2=-\frac{a_{21}^2}{12}$. The
first blow-up along the shifted singularity is again
$X\xleftarrow{(x,y,z|e_1)}X'$. The second blow-up along the shifted
singularity is $X'\xleftarrow{(x,y,e_1|e_2)}X''$. There are no more
codimension 1 singularities after the blow-up. The fibers of $e_1=0$
give the exterior nodes on the $A_4$ Dynkin diagram, and the fibers of
$e_2=0$ give the interior nodes.  The exterior nodes split if and only
if $a_{20}$ is a perfect square, as do the interior nodes. Thus if
$a_{20}$ is a perfect square, $G=SU(5)$; if not, $G=Sp(2)$.

$\mathbf{I_1^*.}$
For $I_1^*$ $f$ series starts with $f_2$ term whereas $g$ series starts with $g_3$ terms. Let $f_2=-3a_{21}^2, g_3=2a_{21}^3$. Blow up the singularity at $(0,0,0)$ by $X\xleftarrow{(x,y,z|e_1)}X'$. Two singular loci appear after this blow-up. We shift the variety to make one of the singularities sit at $x=0, y=0, e_1=0$ then blow it up by $X'\xleftarrow{(x,y,e_1|e_2)}X''$. Two singular locus appear after this blow up and one of them is at $e_1=0, e_2=0, y=0$ and we blow it up by $X''\xleftarrow{(e_1,e_2,y|e_3)}X'''$. We are left with one singularity after this blow-up and this singularity is given by an algebraic equation of the form $e_2A = yB$. After a small resolution $X'''\xleftarrow{(e_1,y|e_4)}X''''$ there are no more codimension 1 singularities. There are four divisors when we go into the exceptional locus. We check the case when they split. The split condition is given by letting $f_3=c_1a_{21}a_{11}^2, g_4=c_2a_{21}^2a_{11}^2$, $c_1, c_2$ are two arbitrary numbers. Non-split case gives us $SO(9)$ and split case gives us $SO(10)$.

$\mathbf{II.}$
Tune $f_0$ and $g_0$ to zero to get a type $II$ fiber. This has no gauge algebra.

$\mathbf{III.}$
Tune $f_0, g_0$ and $g_1$ to zero. $X$ is singular and we blow it up by $X\xleftarrow{(x,y,z|e_1)}X'$. There are no more codimension 1 singularities after the blow-up. The exceptional divisor is fibered by a genus 0 curve which gives us $SU(2)$.

$\mathbf{IV.}$ Tune $f_0, f_1, g_0,$ and $g_1$ to zero. There is a
singularity in the fibration and we blow it up by
$X\xleftarrow{(x,y,z|e_1)}X'$. There are no more codimension 1
singularities after the blow-up. The fiber of the exceptional divisor
splits when $g_2$ is a perfect square, $g_2=a_{31}^2$. In that case $G=SU(3)$; if $g_2$ is not
a perfect square, $G=Sp(1)$.

$\mathbf{I_0^*.}$ Let $f=f_2z^2, g=g_3z^3$. First we do
$X\xleftarrow{(x,y,z|e_1)}X'$. The resulting variety has the form
$y^2=e_1P$ in the ambient space so that we can do a small resolution
$X'\xleftarrow{(e_1,y|e_2)}$. The resulting variety, without any
codimension 1 singularities, when it splits, gives 4 four nodes. The
splitting is given by the relationship
$f_2=-a_{21}^2-a_{21}a'_{21}-a_{21}^{'2},
g_3=a_{21}^2a'_{21}+a_{21}a_{21}^{'2}$. Let's call the node given by
$e_1=0$ $N_1$ and by $e_2=0$ $N_{21}, N_{22}. N_{23}$. $N_1$
intersects with $N_{21}$ $N_{22}$ and $N_{23}$ at 3 different
points. There are no other intersections. This gives us
$SO(8)$. There are two different situations if $I_0^*$ does not fully split. When $f_2$ and $g_3$ are two generic sections, there are 2 nodes given by $e_1=0$ and $e_2=0$, call them $N_1$ and $N_2$ respectively. $N_1$ is fibered by $\mathbb{P}^1$ and $N_2$ is fibered by 3 $\mathbb{P}^1$'s. The 3 $\mathbb{P}^1$'s in $N_2$ are getting mapped to each other when encircling $e_2=0$. The intersection diagram gives us $G_2$. When $f_2=a_{42}-a_{21}^2, g_3=-a_{42}a_{21}$, two of the three nodes $N_{21}$ $N_{22}$ $N_{23}$, say $N_{22}$ and $N_{23}$, when encircling $e_2=0$, get mapped to each other. This intersection diagram gives us $SO(7)$.

$\mathbf{IV^*.}$
Let $f=f_3z^3, g=g_4z^4$. We need to do 4 blow-ups to resolve the singularities until there are no codimension 1 singularities. The exceptional locus split when $g_4=a_{32}^2$, let's call the node given by $e_1=0$ $N_1$, by $e_2=0$ $N_{21}$ and $N_{22}$, by $e_3=0$ $N_{31}$ and $N_{32}$ and by $e_4=0$ $N_4$. $N_1$ intersects with $N_4$, $N_4$ with $N_{31}$ and $N_{32}$ at two different points, $N_{31}$ with $N_{21}$ and $N_{32}$ with $N_{22}$. There are no more intersections. This gives us $E_6$. When $g_4$ is a generic section, $e_2=0$and $e_3=0$ do not split. $N_21$ gets mapped to $N_22$ while encircling $e_2=0$ and $N_31$ to $N_32$ while encircling $e_3=0$. The intersection diagram gives us $F_4$.

$\mathbf{III^*.}$
Let $f=f_3z^3$, $g=g_5z^5$.
We do not perform the blowup explicitly, resolutions yield exceptional divisors that are fibered by genus $0$ curves whose intersection diagram gives us $E_7$. 

$\mathbf{II^*.}$
Let $f=f_4z^4$, $g=g_5z^5$.
We do not perform the blowup explicitly, resolutions yield exceptional divisors that are fibered by genus $0$ curves whose intersection diagram gives us $E_8$. 

\begin{table}[!htbp]
\centering
\begin{tabular}{ |c|c|c|c| }
\hline
Type & Tuned off & Tuned on & Gauge group\\
\hline
$I_2$ & $f_{0}$ $g_{0}$ $g_{1}$ & $a_{20}$ & $SU(2)$ \\
\hline
$I_{3ns}$ & $f_{0}$ $f_1$ $g_{0}$ $g_{1}$ $g_{2}$ & $a_{20}$ $a_{21}$ & $Sp(1)$ \\
\hline
$I_{3s}$ & $f_{0}$ $f_{1}$ $g_{0}$ $g_{1}$ $g_{2}$ & $a_{10}$ $a_{21}$ & $SU(3)$ \\
\hline
$I_{4ns}$ & $f_{0}$ $f_{1}$ $g_{0}$ $g_{1}$ $g_{2}$ $g_{3}$ & $a_{20}$ $a_{21}$ & $Sp(2)$ \\
\hline
$I_{4s}$ & $f_{0}$ $f_{1}$ $g_{0}$ $g_{1}$ $g_{2}$ $g_{3}$ & $a_{10}$ $a_{21}$ & $SU(4)$ \\
\hline
$I_{5ns}$ & $f_{0}$ $f_{1}$ $f_{2}$ $g_{0}$ $g_{1}$ $g_{2}$ $g_{3}$ $g_{4}$ & $a_{20}$ $a_{21}$ & $Sp(2)$ \\
\hline
$I_{5s}$ & $f_{0}$ $f_{1}$ $f_{2}$ $g_{0}$ $g_{1}$ $g_{2}$ $g_{3}$ $g_{4}$ & $a_{10}$ $a_{21}$ & $SU(5)$ \\
\hline
$I_{1ns}^*$ & $f_{0}$ $f_{1}$ $f_{2}$ $g_{0}$ $g_{1}$ $g_{2}$ $g_{3}$ & $a_{21}$ & $SO(9)$ \\
\hline
$I_{1s}^*$ & $f_{0}$ $f_{1}$ $f_{2}$ $f_{3}$ $g_{0}$ $g_{1}$ $g_{2}$ $g_{3}$ $g_{4}$ & $a_{21}$ $a_{11}$ & $SO(10)$ \\
\hline
$II$ & $f_{0}$ $g_{0} $ & -- & -- \\
\hline
$III$ & $f_{0}$ $g_{0}$ $g_{1}$ & -- & $SU(2)$ \\
\hline
$IV_{ns}$ & $f_{0}$ $f_{1}$ $g_{0}$ $g_{1}$ & -- & $Sp(1)$ \\
\hline
$IV_s$ & $f_{0}$ $f_{1}$ $g_{0}$ $g_{1}$ $g_{2}$ & $a_{31}$ & $SU(3)$ \\
\hline
$I_{0ns}^*$ & $f_{0}$ $f_{1}$ $g_{0}$ $g_{1}$ $g_{2}$ & -- & $G_2$ \\
\hline
$I_{0s_1}^*$ & $f_{0}$ $f_{1}$ $f_{2}$ $g_{0}$ $g_{1}$ $g_{2}$ $g_{3}$ & $a_{21}$ $a_{42}$ & $SO(7)$ \\
\hline
$I_{0s_2}^*$ & $f_{0}$ $f_{1}$ $f_{2}$ $g_{0}$ $g_{1}$ $g_{2}$ $g_{3}$ & $a_{21}$ $a'_{21}$ & $SO(8)$ \\
\hline
$IV_{ns}^*$ & $f_{0}$ $f_{1}$ $f_{2}$ $g_{0}$ $g_{1}$ $g_{2}$ $g_{3}$ & -- & $F_4$ \\
\hline
$IV_s^*$ & $f_{0}$ $f_{1}$ $f_{2}$ $g_{0}$ $g_{1}$ $g_{2}$ $g_{3}$ $g_{4}$ & $a_{32}$ & $E_6$ \\
\hline
$III^*$ & $f_{0}$ $f_{1}$ $f_{2}$ $g_{0}$ $g_{1}$ $g_{2}$ $g_{3}$ $g_{4}$ & -- & $E_7$ \\
\hline
$II^*$ & $f_{0}$ $f_{1}$ $f_{2}$ $f_{3}$ $g_{0}$ $g_{1}$ $g_{2}$ $g_{3}$ $g_{4}$ & -- & $E_8$ \\
\hline
\end{tabular}
\caption{Sections needed to be tune off or on to get the required gauge group. $f_i\in\cO(-4K_Z+(4-i)N_{Z|B}), g_i\in\cO(-6K_Z+(6-i)N_{Z|B}), a_{ni}\in\cO(-nK_Z+(n-i)N_{Z|B})$. }
\label{appendixtable1}
\end{table}

\bibliography{refs}

\end{document}